\documentclass[english,a4paper,11pt]{article}
\pdfoutput=1
\usepackage{amssymb, latexsym, amsmath, bbm, color}
\usepackage{hyperref}
\usepackage{amsfonts}
\usepackage{graphicx}
\usepackage[english]{babel}
\usepackage{natbib}
\DeclareMathOperator{\Tr}{Tr}
\title{The static quark potential from a multilevel algorithm for the improved gauge action}
\author{\large{Anne Mykk\"anen} \\
\\
  \multicolumn{1}{p{.7\textwidth}}{\centering{\small{Department of Physics and Helsinki Institute of Physics, \\P.O. Box 64, FI-00014 University of Helsinki, Finland\\email: anne-mari.mykkanen@helsinki.fi}}}}
\date{}

\begin{document}

\title{\begin{flushright}
       \mbox{\normalsize HIP-2012-18/TH}
       \end{flushright}
       \vskip 20pt
The static quark potential from a multilevel algorithm for the improved gauge action}

\maketitle

\begin{abstract}
We generalize the multilevel algorithm of L\"uscher and Weisz to study SU($N$) Yang-Mills theories with the tree-level improved gauge action. We test
this algorithm, comparing its results with those obtained using the Wilson action, in SU($3$) and SU($4$) Yang-Mills theories in $2+1$ and $3+1$ dimensions. We measure the static quark potential and extract the L\"uscher term, predicted by the bosonic string theory.
\end{abstract}

\section{Introduction}

An efficient lattice gauge theory algorithm proposed by L\"uscher and Weisz, the multilevel algorithm, has been shown to provide an exponential reduction of the statistical errors in calculations of the Polyakov loop correlation function \citep{LW6}. Multilevel algorithms are useful in many contexts, in which one has to cope with an exponentially decaying signal-to-noise ratio, for example in the computation of the glueball spectrum \citep{Caselle26} and in the computation of the correlation functions, which are related to the transport coefficients of the quark-gluon plasma \citep{added10, added11}. In \citep{addedLaineMeyer}, a multilevel algorithm was used to study the localization properties of gauge fields on domain wall defects. In this paper, we generalize the multilevel algorithm, and, as an example of application, we use it to study the static quark potential.

In a confining theory, the static quark potential has the large distance asymptotic expansion \citep{LuscherWeisz}
\begin{equation}
V(r)=\sigma r +\mu +\gamma /r +O(1/r^2),
\label{OG_potential}
\end{equation}
where $\sigma$ is the string tension, $\mu$ a constant (a regularization-dependent mass), and $\gamma$ is the L\"uscher term
\begin{equation}
\gamma=-\frac{\pi}{24}(D-2)
\label{luscher_term}
\end{equation}
with $D$ as the dimension of the space-time \citep{LW2, LW3}. The interaction between the static quark-antiquark pair can be described by an effective string theory. It has been suggested that the expectation values of large Wilson loops have a correspondence with amplitudes of an effective bosonic string theory \citep{LuscherWeisz, Nambu}. The Polyakov loop correlator can be used similarly to the Wilson loop to study string effects \citep{LW7, marco31, LW8, LW9}.

One of the consequences of the effective string description at zero temperature is the presence of the term proportional to $1/r$, i.e. the L\"uscher term (\ref{luscher_term}), in the long distance inter-quark potential. The L\"uscher term includes a coefficient that depends only on the dimension of the space-time and is not influenced by higher order corrections of the effective string action. The aim of this paper is to study the potential, namely compute the L\"uscher term, in pure glue SU($N$) Yang-Mills theory, using numerical simulations.

Although the validity of the asymptotic expansion (\ref{OG_potential}) can be checked by means of numerical simulations, the problem in these lattice computations, is that the signal-to-noise ratio decays quickly at large distances. This makes it difficult to clearly separate the $\gamma/r$ correction from the other terms. Still, lattice gauge theory can offer support for the string model. For example, the data for the potential in a range of distances from $0.4$ to $0.8$ fm has been noted to agree with (\ref{OG_potential}) within small errors \citep{LW4}. Also, using highly efficient simulation techniques it was confirmed \citep{LW5} that the expectation values of large rectangular Wilson loops are matched by the string theory amplitudes to a precision where the subleading string effects can be observed. However, a more recent study \citep{wilson_loops} on string effects at large-$N$ using Wilson loops seemed to differ from the effective string prediction, contrary to previous studies \citep{Teper1, Teper2}.  

In this paper, all simulations are performed utilizing the multilevel algorithm, and besides the standard Wilson action, we will also conduct simulations with the tree level improved action \citep{added1, added2, added3, added4}. The improved action can strongly reduce the lattice artifacts, and hence give better results also on relatively small lattices. Examples of ``success stories" of the improved action for the gauge action include computations of the SU($3$) equation of state \citep{CasimirScaling46, added6} and the studies of renormalized Polyakov loops \citep{added7, added8}. A high-precision determination of the scale with the L\"uscher-Weisz improved action for the SU($3$) Yang-Mills theory was carried out in \citep{addedGattringer}. One of the goals of the present work consists in showing that the multilevel algorithm allows one to obtain a similar level of precision, with a more limited computational effort. Similarly, improved actions for the quarks are particularly useful when it comes to computationally challenging problems, like, e.g. lattice studies of walking technicolor models (see, for example, \citep{addedtechni}). 

The string-like features of the interquark `flux tube' have already been studied with calculations in several gauge models \citep{aharony26}, including $\mathbb{Z}_2$ \citep{LW5, marco19, marco30, Rajantie1210.1106}, $\mathbb{Z}_4$ \citep{Z4}, U($1$) \citep{Marco0503024, marcoU1, Koma}, SU($2$) \citep{Caselle, marco22, marco30, Bonati}, SU($3$) \citep{LuscherWeisz, marco28, Bali15}, as well as with SU($N > 3$) \citep{Teper1, Teper2, aharony27, aharony28}, (see \citep{TepersReview, Lucini1210.4997, Marco1210.5510} for reviews), $\mathrm{G}_2$ gauge theory \citep{addedG2}, and even in a random percolation model (with an appropriate definition of the observables) \citep{Gliozzi:2005ny, aharony33}. By extending the research also to bigger gauge groups we pursue two main subjects of interest. First of all, since large-$N$ studies are largely motivated by the fact that they can provide a mathematically simplified frame for studying QCD \citep{addedtHooft}, we want to see whether the numerical results for the flux tube obtained with a large-$N$ theory agree with SU($3$) results. A second motivation comes from the conjectured connection between large-$N$ conformal gauge theories and string theory (see for example the reviews \citep{Aharony:1999ti} and \citep{MateosString}), affiliating with the Maldacena conjecture and AdS/CFT. In this paper, the groups SU($3$) and SU($4$) are studied.

This paper is organized as follows: In section $2$, as an extended introduction on the topic of this paper, we will discuss some basic issues on bosonic string theory, namely the Nambu-Goto string and how it can model the behavior of the flux tube in a confining gauge theory, and also some earlier lattice results relevant to this topic. Then, in section $3$ we discuss the Polyakov loop correlation function on the lattice and the actions used in this work, and in section $4$ we describe the multilevel algorithm in more detail. In sections $5$ and $6$ we present the results of this work and the conclusions.

\section{Bosonic string theory}

\subsection{Nambu-Goto string}

The simplest effective action for a bosonic string is simply the string tension $\sigma$ times the area of the string world sheet, i.e. the Nambu-Goto action \citep{marco35, marco36, marco37} 
\begin{equation}
S_{\mathrm{eff}}=\sigma \int d^2 \xi \sqrt{\textrm{det} g_{\alpha \beta}}.
\end{equation}
In the so called ``physical gauge'' the integrand reads
\begin{equation}
\sqrt{ 1 + (\partial_0 h)^2 + (\partial_1 h)^2 + (\partial_0 h \times \partial_1 h)^2 }
\end{equation}
where $h$ is the displacement of the world sheet surface in the transverse directions. Expanding this in a perturbative series in $1/(\sigma r^2)$, at leading order we get the following expression for the Polyakov loop correlation function (see, e.g. \citep{Marco0503024} for a discussion)
\begin{equation}
\langle P^{\star}(r)P(0) \rangle = \frac{e^{-\sigma rL-\mu L}}{\Big(\eta\Big(i\frac{L}{2r}\Big)\Big)^{D-2}},
\label{pol_expect}
\end{equation}
where we have used Dedekind's $\eta$ function
\begin{equation}
\eta(\tau)=q^{\frac{1}{24}}\prod_{n=1}^{\infty}(1-q^n), \qquad q=e^{2\pi i\tau}.
\end{equation}
When $\frac{L}{2r} \gg 1$, eq. (\ref{pol_expect}) gives the L\"uscher term in the quark-antiquark potential.

The spectrum of the Nambu-Goto string can be obtained by canonical quantization \citep{marco39, marco40}: the energy levels for a string with fixed ends are
\begin{eqnarray}
E_n(r) &=& \sigma r \sqrt{1+\frac{2\pi}{\sigma r^2}\Big( n-\frac{D-2}{24} \Big)}, \qquad n \in \mathbb{N}, \\
&=& \sigma r - \frac{(D-2-24n)\pi}{24r}+O(1/r^3).
\end{eqnarray}
As a consequence, the partition function describing the string with fixed ends reads
\begin{equation}
Z=\sum^{\infty}_{n=0}\omega_n e^{-E_n(r)L}
\end{equation}
where $\omega_n$ are the number of states. 

As mentioned, one of the outcomes of the effective string description at zero temperature is the presence of the L\"uscher term in the long distance inter-quark potential. One should not confuse this with a Coulomb term originating from the one-gluon exchange process. The L\"uscher term is a Casimir effect, which is due to the finiteness of the interquark distance $r$. As a heuristic argument for the term, let us consider the string-like tube created by two color sources separated by distance $r$. Due to quantum fluctuations, the string can vibrate, and we can express its wavelength through 
\begin{equation}
r=\lambda \Big( \frac{1+n}{2} \Big)
\end{equation}
where $n=0,1,2,3,\dots$. If we consider each of these modes as quantum mechanical harmonic oscillators, then their energies are
\begin{equation}
E_k^n=\hbar \omega_n \Big(k+\frac{1}{2}\Big)
\end{equation}
and if we just consider the ground state of the system, $k=0$, we can write its energy $E_0$ as
\begin{eqnarray}
E_0 &=& \sum_{n=0}^{\infty}\frac{1}{2}\hbar \omega_n =\frac{1}{2}\sum_{n=0}^{\infty}\frac{2\pi}{\lambda_n}=\pi \sum_{n=0}^{\infty}\frac{1}{\lambda_n} \nonumber \\
&=& \frac{\pi}{r} \Big(\frac{1}{2}+1+ \frac{3}{2} +2+ \frac{5}{2}+\dots \Big) \nonumber \\
&=& \frac{\pi}{2r}\Big(1+2+3+4+5+\dots \Big) \nonumber \\
&=& -\frac{1}{2}\frac{\pi}{12r}
\end{eqnarray}
where we have used the Riemann zeta regularization. This is just for one transverse direction; in $D$ dimensions there are $D-2$ transverse directions, so in four dimensions, we have
\begin{equation}
E_0=-\frac{\pi}{12r}.
\end{equation}
Finally, the effective string model also gives a prediction for the form of the inter-quark potential in finite temperature gauge theories \citep{Bonati, Caselle, bonati7}.

\subsection{QCD string}
In the confining regime of SU($N$) gauge theories, the asymptotic behavior of $V(r)$ at large distances is a linear rise, and the flux lines between the well separated color sources are expected to be squeezed in a thin, string-like tube \citep{LW2, LW3, marco5}. In a pure gauge theory the ground state interquark potential $V(r)$ of a heavy $Q\bar{Q}$ pair can be expressed in terms of the two-point correlation function $G(r)$ of Polyakov lines
\begin{equation}
V(r)=-\frac{1}{L}\log G(r)=-\frac{1}{L} \log \langle P^{\star}(r)P(0) \rangle,
\label{pot_corr}
\end{equation}
where $r$ is the interquark distance and $L$ the system size in the time-like direction \citep{Marco0503024}.

Assuming that the low energy dynamics of the pure gauge model is described by the effective string, and assuming that decays of excited states do not play a significant role, we can write the Polyakov loop two-point correlation function as a string partition function 
\begin{equation}
\langle P^{\star}(r)P(0) \rangle =\int \mathcal{D}h e^{-S_{\mathrm{eff}}},
\end{equation}
where $S_{\mathrm{eff}}$ is the effective action for the world sheet spanned by the string.

The dynamics of the confining string is not known, but the general properties of the effective string can be derived based on the symmetries it should have \citep{Gliozzi, HBMeyer, AharonyField, Dodelson, Klinghoffer}. Since the confining string should respect the expected rotational symmetries, only the rotationally symmetric terms can be considered part of the effective string action. In fact these constraints are more general and they restrict the form of the effective action (at least at the lowest orders in $1/(\sigma r^2)$) to be the Nambu-Goto one. Monte Carlo computations on a lattice may offer a way to study the low energy effective action of a confining string and give insights of its properties. Lattice simulations for pure Yang-Mills theories in $D=3$ and $D=4$ show the effective action being very well approximated by this form. However, deviations from Nambu-Goto can be derived at higher orders (see \citep{Brandt} and references therein).

Different approaches allow one to constrain the effective action of a confining string. Polchinski and Strominger \citep{aharony2} proposed to consider the degrees of freedom in the effective action as the embedding coordinates of the string in a conformal gauge world sheet (see also \citep{Drummond:2004yp, HariDass:2006sd, Drummond:2006su, Dass:2006ud, HariDass:2007gn}). Requiring the effective action to have the correct critical central charge, one is left with constraints that have been shown to imply the four-derivative effective action to agree with the Nambu-Goto form. Unfortunately, generalizing this approach to higher orders appears to be challenging.

Another approach \citep{LW2, LW3, aharony7} is to write the effective action in static gauge, such that the degrees of freedom are only the $(D-2)$ transverse fluctuations of the string world sheet. Essentially, the action should non-linearly realize the Lorentz symmetry rotating the direction in which the string propagates, and the transverse directions. Following the suggestion of L\"uscher and Weisz in \citep{aharony7}, the effective action is constrained by computing the partition function of long closed strings, winding around a periodic size of the system, and writing it in terms of a sum over string states. 

The seminal reference \citep{aharony7} generated a fruitful line of research: in particular the method was extended in \citep{Aharony} to any $D$ and to the closed strings. In \citep{AharonyField} it was apprehended, that the Lorentz symmetry of the underlying Yang-Mills theory had a very important role in the L\"uscher-Weisz argument, and in fact, the whole Nambu-Goto action was necessary to respect the Lorentz symmetry. Aharony and Dodelson first suggested a bulk correction of the Nambu-Goto action in \citep{Dodelson}. This, however, was critically discussed by Dubovsky et al. in \citep{Dubovsky1203.1054} and later, in the ref. \citep{Gliozzi}, a general construction of the Lorentz invariant bulk action was formulated. In principle the effective action could also contain boundary terms, which have been discussed by Bill\'o et al. in \citep{Billo1202.1984}. It is important to remark that, as shown in ref. \citep{AharonyField}, these terms are compatible with Lorentz invariance. These boundary terms were compared with numerical results of simulations of $3D$ $\mathbb{Z}_2$ gauge theory in \citep{Billo1202.1984} and in $3D$ SU($2$) gauge theory in \citep{Brandt}.

Even though the interquark potential $V(r)$ and related quantities have been studied extensively on the lattice for many years (see, for example, \citep{Bali15, Bali16, aharony26} for references), the question whether the picture given by effective string models is quantitatively satisfactory, is still under investigation. Previous results \citep{Bali15, Bali16} have shown prominent support for the bosonic string prediction, in particular, the L\"uscher term has been shown to be a universal feature of the IR regime of confined gauge theories. However, results from recent, high precision Monte Carlo simulations \citep{Koma, marco17, LuscherWeisz, marco19, Caselle12, marco22, marco23, Caselle, marco25, marco26} suggest that higher order corrections to $V(r)$ might not be universal. In fact, there are theoretical arguments suggesting that the effective string action is different from the Nambu-Goto one at higher orders.

Another aspect of interest, is the excitation spectrum description. According to the bosonic string picture, the excitations are expected to be described by the spectrum of harmonic oscillator energies
\begin{equation}
E_k=E_0+\frac{\pi}{r}k, \qquad k\in \mathbb{N}.
\end{equation}
In the ground state potential the L\"uscher term is supported by numerical evidence down to very short distances, but for the excited states the lattice results are not in such a good agreement with the theoretical expectations, as discussed in \citep{Marco0503024}. Mismatches between effective string predictions \citep{LW7} and numerical results have been found in \citep{marco27, marco28, marco29, marco30}.

\subsection{Some earlier lattice results}

In recent years the effective action of confining strings has been studied on the lattice with higher and higher precision (see \citep{aharony26, TepersReview, Lucini1210.4997, Marco1210.5510} for reviews of results). As discussed in \citep{Aharony}, studies of the three-dimensional pure Yang-Mills theory have produced very accurate results of the spectrum of confining flux tubes in large-$N$ SU($N$) gauge theories \citep{aharony27, aharony28}. Considering torelons of length $L$, it has been found that the ground state energy agrees with the Nambu-Goto result at order $1/L$, and is consistent with it at order $1/L^3$. Also, a possible deviation at order $1/L^5$ comes with a very small coefficient. The excitation spectrum has also been found to agree with Nambu-Goto at orders $1/L$ and $1/L^3$, however at higher orders there are deviations. The lattice data are now reaching high precision, so that it is becoming possible to determine at which order the deviation occurs, which could be already at order $1/L^5$ \citep{Aharony, Gliozzi}. 

Similarly, in simulations of large-$N$ gauge theories done in $3+1$ dimensions, there is agreement with Nambu-Goto for large $L$, but the order in which the deviations arise is not clear. Such is the case also in simulations of interfaces in the $2+1$-dimensional Ising model, in which one can also see a good agreement with Nambu-Goto \citep{aharony29}, but still more precision would be needed in order to tell at what order deviations from Nambu-Goto arise \citep{Aharony}. 

$2+1$ dimensional confining strings in higher representations, called ``$k$-strings", have also been studied \citep{aharony30, aharony31} at large $N$. Comparison to Nambu-Goto showed large deviations for all states, including the ground state, possibly starting already at order $1/L^5$. However, there are some technical aspects to be taken into account. In the large-$N$ limit, assuming $k$ to be fixed, the binding energy of $k$ fundamental strings to form a $k$-string may vanish as $1/N^2$ or as $1/N$ \citep{aharony32}, implying that in the large-$N$ limit there are at least $(k-1)(D-2)$ light modes on the worldvolume of a $k$-string, whose mass goes to zero in the large-$N$ limit as $1/N$ or as $1/\sqrt{N}$, respectively. The relevant length scales concerning the form of the effective action are larger than $N/\sqrt{T}$ (or $\sqrt{N/T}$).

Finally, we mention that the ground state potential of a confining string in the continuum limit of a percolation model in $2+1$ dimensions was studied in \citep{aharony33}, and again agreement was found with Nambu-Goto at orders $1/L$ and $1/L^3$, however deviations were seen at order $1/L^5$. As discussed in \citep{Aharony}, this model is not necessarily expected to correspond to a weakly coupled string theory, but it is interesting in itself.

\section{Polyakov loop correlation function on the lattice}

We discretize the SU($N$) Yang-Mills theory on an isotropic $4$-dimensional, or alternatively, $3$-dimensional lattice with spacing $a$, time-like extent $aN_t$ and spatial size $aN_s$. Apart from providing precise results more quickly, simulations in $3$ dimensions are interesting because there the $1/r$ term is certain to be the L\"uscher term; in $4D$ it could get mixed with a Coulomb interaction. In $3D$ the Coulomb interaction is of the form $\log(r)$, rather than $1/r$. 

In all directions of the lattice we impose periodic boundary conditions. For the gauge field action we take the usual Wilson plaquette action
\begin{equation}
S=\beta \sum_{x,\mu < \nu} \Big(1-\frac{1}{N}\textrm{Re} \Tr U_{\mu\nu}(x)\Big),
\label{wilson_action}
\end{equation}
where $U_{\mu\nu}$ is the plaquette oriented along the $\mu\nu$-plane and located at the lattice site $x$. In $3+1$ dimensions $\beta=2N/g^2$ and in $2+1$ $\beta=2N/(ag^2)$. We also carry out the simulations using the tree-level improved action \citep{added1, added2, added3, added4}
\begin{equation}
S = \beta \sum_{x, \mu < \nu} \left\{ 1 - \frac{1}{N} \textrm{Re} \Tr \left[ \frac{5}{3} U^{1,1}_{\mu\nu}(x) - \frac{1}{12} U^{1,2}_{\mu\nu}(x) - \frac{1}{12} U^{1,2}_{\nu\mu}(x) \right] \right\} .
\label{improved_action}
\end{equation}

For any given gauge field configuration $U(x,\mu)$, the (trace of the) Polyakov loop is defined as
\begin{equation}
P(x) = \Tr{U(x,\mu)U(x + \hat{a}\mu, \mu) \dots U(x + (T-a)\hat{\mu},\mu)}_{\mu=0}
\end{equation}
and the correlation function
\begin{equation}
\langle P(x)^{\star}P(y) \rangle = \frac{1}{Z}\int \prod_{x,\mu}dU(x,\mu) P(x)^{\star}P(y)e^{-S[U]}.
\end{equation}
The ground state inter-quark potential $V(r)$ of a heavy $\bar{Q}Q$ pair in a pure gauge theory can be expressed with the two-point correlators of the Polyakov loop \citep{LuscherWeisz}
\begin{equation}
V(r)=-\frac{1}{aN_t}\log\langle P(x)^{\star}P(y) \rangle +\epsilon,
\label{potential}
\end{equation}
where
\begin{equation}
\epsilon=\frac{1}{aN_t}(\omega_1 e^{-\Delta ET}+\dots),\qquad \Delta E=E_1-E_0.
\end{equation}

\section{The Multilevel algorithm}

The lattice data in this work is obtained implementing the multilevel algorithm of L\"uscher and Weisz \citep{LW6}. In this algorithm, using the locality of the theory, the lattice is split into sublattices that do not communicate with one another, and the final observables are built combining the independent measurements of each sublattice. Of course, from time to time the boundaries between the sublattices are updated, so that the final results are the same as in the usual theory.

Following the discussion in \citep{Caselle}, consider an observable $\mathcal{O}$, calculated by combining the results of averages $\mathcal{O}_{sub}$ computed in $\mathcal{N}$ different sublattices using the Wilson action. With $N_{meas}$ sublattice measurements, the combination of the sublattice averages $\mathcal{O}_{sub}$ corresponds to $(N_{meas})^\mathcal{N}$ measurements of $\mathcal{O}$, i.e. we get an estimate of $\mathcal{O}$ as if $(N_{meas})^\mathcal{N}$ measurements would have been performed. Due to the links that have been kept frozen at the boundaries of the sublattices, this estimate is biased by a background field, but this bias is removed by averaging over the boundary configurations.

Using the notation of \citep{Caselle}, consider the correlation function of two Polyakov loops
\begin{eqnarray}
\langle P(\vec{0}) P(\vec{x})^* \rangle & = & \frac{1}{Z} \int \prod_{y,\mu}dU_{y,\mu}  \Tr \Big[ U_{(\vec{0},0),t} \dots U_{(\vec{0},L-1),t} \Big] \nonumber \\ 
& \times & \Tr \Big[ U_{(\vec{x},0),t}^* \dots U_{(\vec{x},L-1),t}^* \Big] e^{-S[U]} .
\label{ml_pol_corr}
\end{eqnarray}
Next, we slice the lattice along the temporal direction into $\mathcal{N}=N_t/n_t$ sublattices, i.e. $n_t$ is the temporal thickness of each sublattice in units of the lattice spacing $a$. Now, to obtain sublattices isolated from each other, we fix the set $V_k^s$ of all spatial links with time coordinates $kn_t$, $k=0,\dots,(\mathcal{N}-1)$. This way, the dynamics of every sublattice depends only on the background field of the two frozen time slices, and hence are totally independent from one another.

As in \citep{Caselle}, we rewrite (\ref{ml_pol_corr}) as
\begin{eqnarray}
\label{sliced}
\langle P(\vec{0}) P(\vec{x})^* \rangle &=& \int \prod_k dU_k^{(s)} T^{\alpha \gamma \beta \delta}_{\vec{0}, (\vec{x})}[V_0^{(s)}, V_1^{(s)}] \nonumber \\ 
& \dots & T^{\epsilon \alpha \rho \beta }_{\vec{0}, (\vec{x})}[V_{\mathcal{N}-1}^{(s)}, V_0^{(s)}] \mathcal{P}[V_k^{(s)}]
\end{eqnarray}
where
\begin{eqnarray}
T^{\alpha \gamma \beta \delta}_{\vec{0}, \vec(x)}[V_i^{(s)}, V_j^{(s)}] & \equiv & \int \prod_{y,\mu}dU_{y,\mu} \Big[ U_{(\vec{0},0),t} \dots U_{(\vec{0},n_t-1),t} \Big]_{\alpha \gamma} \nonumber \\ 
& \times & \Big[ U_{(\vec{x},0),t}^* \dots U_{(\vec{x},n_t-1),t}^* \Big]_{\beta \delta} \frac{e^{-S[U;V_i^{(s)},V_j^{(s)}]}}{Z[V_i^{(s)},V_j^{(s)}]}.
\end{eqnarray}

The partition function of the sublattice with fixed temporal boundaries reads
\begin{equation}
Z[V_i^{(s)},V_j^{(s)}] \equiv \int \prod_{y,\mu}dU_{y,\mu}e^{-S[U;V_i^{(s)},V_j^{(s)}]}
\end{equation}
in which $S[U;V_i^{(s)},V_j^{(s)}]$ is the action in the sublattice with fixed temporal boundaries $V_i^{(s)}$ and $V_j^{(s)}$. We denote color indices $\alpha$, $\beta$, $\gamma$ and $\delta$, and $T^{\alpha \gamma \beta \delta}$ are gauge-invariant quantities under sublattice gauge transformations. $\mathcal{P}[V_k^{(s)}]$ is the probability for the spatial links with time coordinates $kn_t$, $k = 0, \dots, (\mathcal{N}-1)$, to be $V_k^{(s)}$, and can be written
\begin{equation}
\label{propab}
\mathcal{P}[V_k^{(s)}]=\frac{1}{Z}\int \prod_{y,\mu}dU_{y,\mu}\prod_k \delta(U_k^{(s)}-V_k^{(s)})e^{-S[U]}.
\end{equation}

Iterating this procedure and averaging the background field according to the probability distribution of eq. (\ref{propab}) gives a numerical estimate for the integral (\ref{sliced}) \citep{Caselle}. 

In the numerical simulations, three parameters have to be fixed: the temporal sublattice thickness $n_t$, the number $N_{meas}$ of sublattice measurements and the number $M$ of background field configurations to integrate over $V^{(s)}$. The optimal values of these parameters are dependent on each other as well as on the temporal extension $N_t$ of the whole lattice and on the distance between the two Polyakov loops. Finding the optimal choice for them is thus not trivial. Some results on the optimization step can be found in \citep{Caselle12, Caselle26}.

When measuring the Polyakov loop correlation function $\langle P(\vec{0}) P(\vec{x})^* \rangle$ the final error bar is the combination of the uncertainties of the sublattice averages and their fluctuations, due to different background fields. With $n_t$ fixed, a large distance between two loops requires both $N_{meas}$ and $M$ to be large. $N_{meas}$ is typically of the order of several thousands and $M$ of few hundreds \citep{Caselle}. It should be noted that $N_{meas}$ does not depend on $N_t$ whereas $M$ does. With the L\"uscher and Weisz algorithm, exponential gain in the precision of the numerical estimation of $\langle P(\vec{0}) P(\vec{x})^* \rangle$ is possible only in the temporal direction. Every sublattice estimate decreases exponentially with the distance between the loops, but it is still estimated with an error reduction proportional to $1/\sqrt{N_{meas}}$ \citep{Caselle}.

\begin{figure}[h!t]
\centering
\includegraphics[width=1\linewidth]{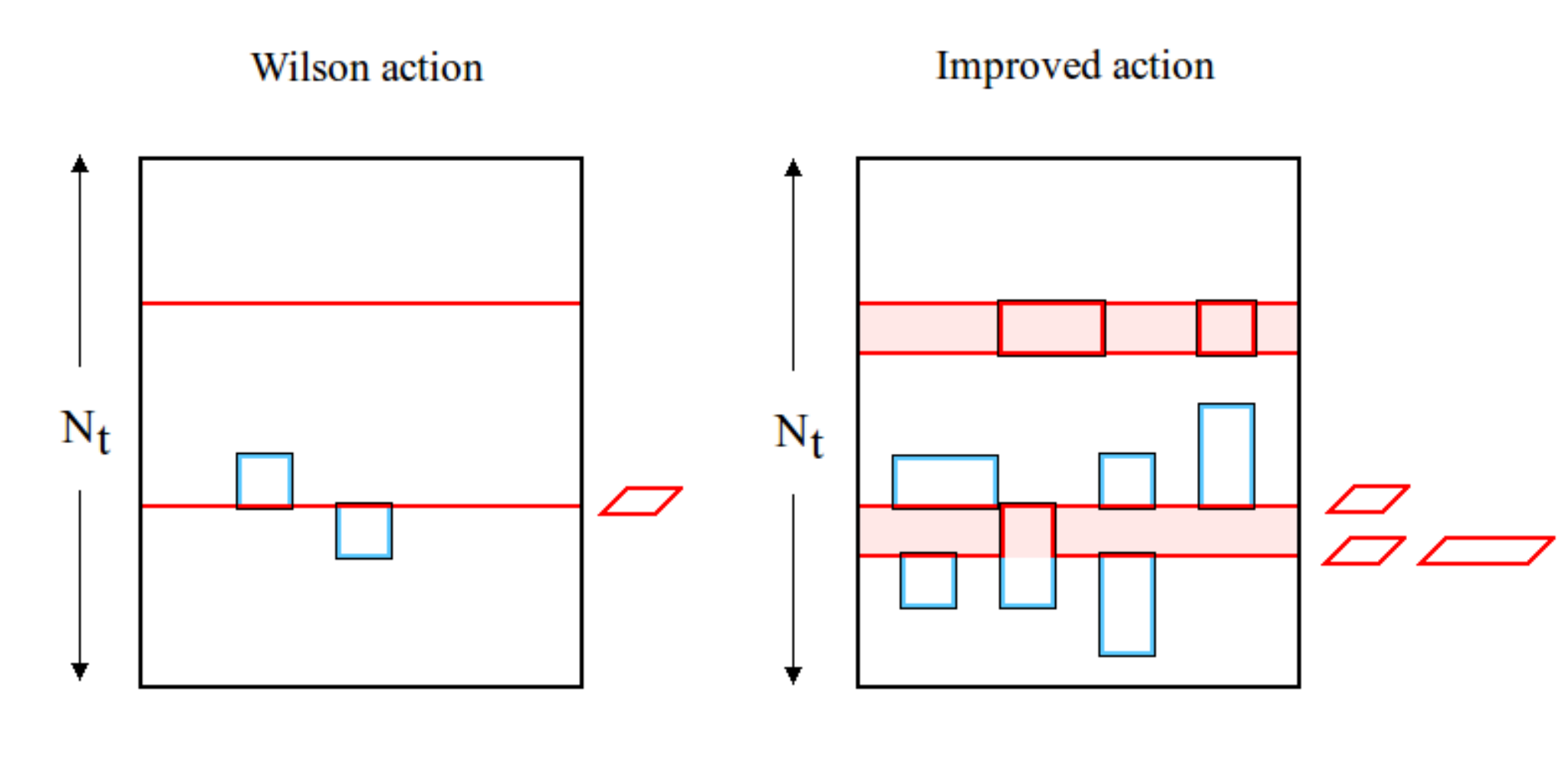}
\caption{For an example on how the multilevel algorithm works, i.e. which links are updated during the sublattice updates, consider a lattice divided into three sublattices. On the left, a presentation of the Wilson action, and on the right, the improved action. On the borders of a sublattice, the red links are the ones kept fixed, whereas the blue links are updated. The improved action is slightly less local than the Wilson action, because it involves not only plaquettes but also rectangles. However, like the Wilson action, the improved action is still ``ultralocal", namely, in the action, each link is coupled directly to other links only up to a finite distance.}
\label{illustration}
\end{figure}

In this work, we generalize this algorithm to the tree-level improved action: since the latter also involves products of link variables around rectangles of size $1 \times 2$, the sublattice averages are done by keeping the spatial links on neighboring time-slices, and the temporal links in between, fixed, as shown in figure \ref{illustration}.

\section{Results}

The simulations were performed with gauge groups SU($3$) and SU($4$), in $2+1$ and $3+1$ dimensions, with the general Wilson action (\ref{wilson_action}) and with the tree-level improved action (\ref{improved_action}). The totalling $8$ different cases (up to $4$ different values of $\beta$ for each) were studied using the multilevel algorithm with sublattice size of $4$ lattice spacings. In $3D$ with SU($3$) the lattice size was $N_t=48$, $N_s=24$, in the other cases $N_t=24$, $N_s=16$. 

The code was running in embarrassingly parallel mode, with 100 nodes. In each of the 100 runs, the first $1000$ updates were used for thermalization, and the following updates, up to $5000$, were used for the actual measurements of the Polyakov correlation function (resulting in a total statistics of up to $5 \times 10^5$ independent, thermalized measurements, for each choice of the parameters).

In addition to considering the correlators measured with an ``on-axis" increasing distance between the sources, we also measure the correlators in ``off-axis" points, i.e. where the sources have ``vertical" distance in addition. In general, the lattice breaks rotational symmetry, so off-axis correlators of Polyakov loops are not expected to lie exactly aligned with the on-axis ones. The full continuum rotational symmetry is however restored
when the lattice spacing tends to zero \citep{LangRebbi}. In the off-axis points one could expect to see a difference between the results obtained using the general Wilson action and the tree-level improved action; when calculating the potential, the off-axis points are expected to fall further from the expected curve with the non-improved action. This makes measuring the off-axis points a valid test to assess the ``goodness" of the improved action.

For each value of $\beta$, the static quark potential was calculated according to (\ref{potential}), see figures \ref{run1}-\ref{run8}. In $4D$ cases, a function of the form
\begin{equation}
V(r)=\sigma r +V_0 +\gamma /r
\label{4dfit}
\end{equation}
was fitted to the data, thus giving estimates for $\sigma$, $V_0$, and $\gamma$. 

In $3D$ cases the fitted function could basically include also a logarithmic term, i.e. the Coulomb term, which would play a role on small distances. However, including this term makes the fit very unstable on larger distances, as seen also in \citep{log_term}. Instead, following the conclusions of \citep{aharony7}, in $3D$ cases we add a $1/r^3$ term in the fitting function, 
\begin{equation}
V(r)=\sigma r +V_0 +\gamma /r + b/r^3.
\label{3dfit}
\end{equation}
Compared to (\ref{4dfit}), this form gives us better a fit and better estimates for the L\"uscher term $\gamma$. 

In $3D$ the bosonic string prediction for the L\"uscher term is $\gamma \simeq 0.1309$, and in $4D$ $\gamma \simeq 0.2618$. As we can see from the tables \ref{wilson_table}, \ref{impr_table}, in both $3D$ and $4D$, the L\"uscher term extracted from our lattice data follows these predictions.

All statistical uncertainties affecting the results were computed
using standard statistical analysis methods. In particular, the error bars on the values
of the average correlators were computed with the jackknife method, while those on the
potential were calculated by Gaussian propagation of the latter.\footnote{This procedure
turns out to be reliable for our data; computing the uncertainties on the potential using
the jackknife method leads to comparable results: the small differences between the
results obtained with the two methods do not have a really significant impact on the
findings of the study, and we include them among the possible sources of systematic
uncertainties that we discuss in subsection~\ref{systematic}}

\subsection{Systematic uncertainties}
\label{systematic}
As always with results from lattice studies, there are some systematic uncertainties that need to be taken into account. Finite volume effects, finite cut-off effects and systematic effects in the choice of the fitting function are all sources of such uncertainties in our calculations. 

The finite volume effects are exponentially suppressed as $e^{-m_G L}$, where $m_G$ is the mass of the lightest glueball, and $L$ the shortest lattice size. $e^{-m_G L}$ is already very small when $m_G L \gtrsim 4$, so what we want is 
\begin{equation}
L \gtrsim \frac{4}{m_G}.
\end{equation}
The lattice sizes chosen in this work meet this requirement.

\begin{figure}[h!]
\centering
\includegraphics[width=0.95\linewidth]{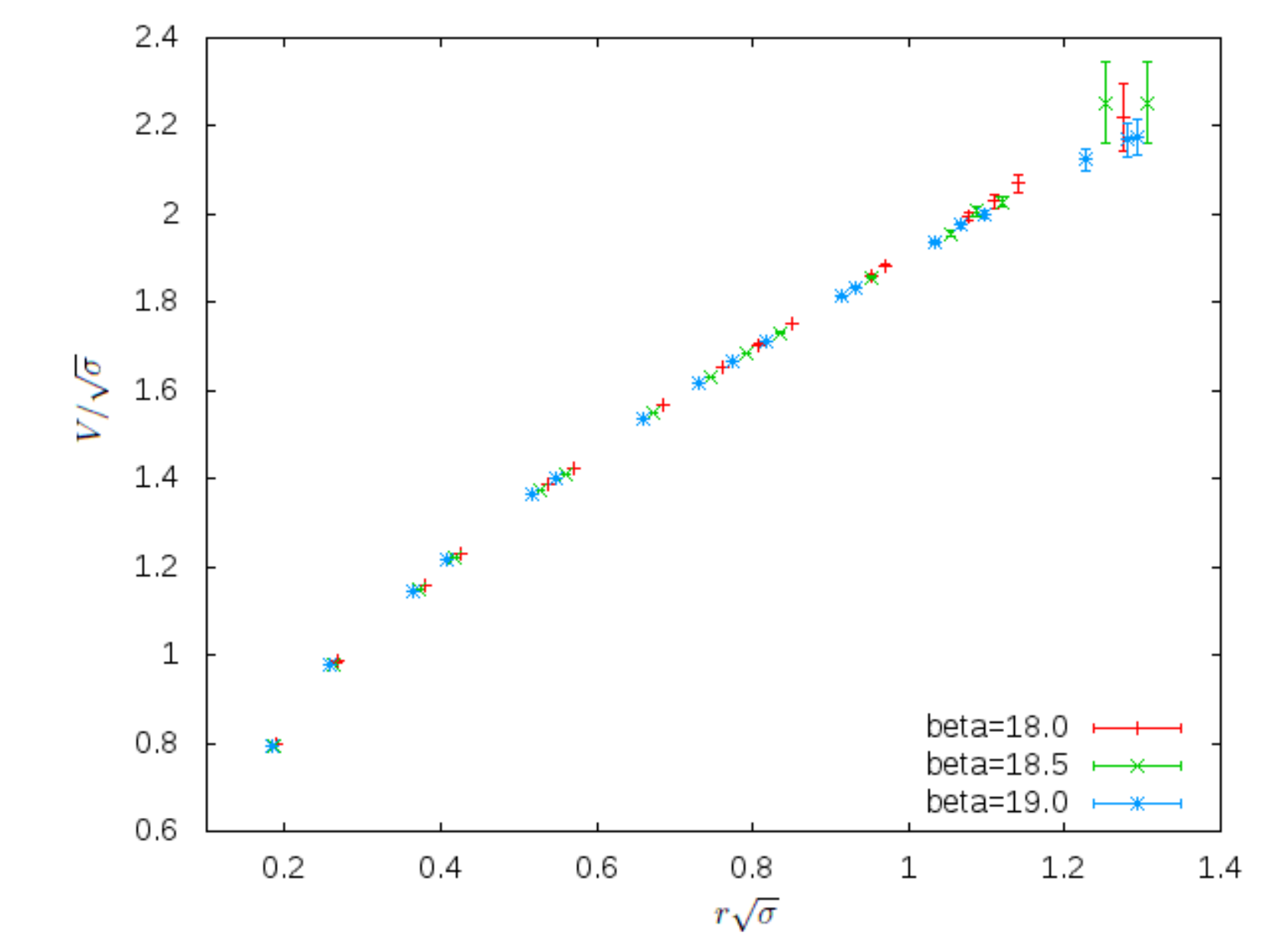}
\caption{The physical values calculated from the results obtained with SU($3$) in $3$ dimensions, with the improved action.}
\label{run2phys}
\end{figure}

As usual, the finite cut-off effects can be controlled by performing simulations at several values of the lattice spacing. The calculations in this paper involve a few different values of $\beta$ in each studied case, and the results indicate the impact from cut-off effects to be very small. As can be seen from figure \ref{run2phys}, the results are clearly already quite close to the continuum, because the points obtained from simulations at different lattice spacing fall on the same curve and because there is no significant difference between points obtained from on-axis versus off-axis correlators (hence the full rotational symmetry is approximately restored).

The systematic effects in the choice of the fitting function can be studied by extending it with an extra term, $c/r^{\alpha}$ with $\alpha=2$ or $3$, and seeing how the fit parameters behave. With SU($3$) in $3D$ and $4D$, adding the extra term with $\alpha=2$ changes $\sigma a^2$ and $V_0 a$ only slightly (for example in $3D$ with $\beta=18.0$, $\sigma a^2$ changes from $0.04482(40)$ to $0.0470(59)$ and $V_0 a$ from $0.2255(23)$ to $0.202(61)$), but gives very poor estimates for $\gamma$. With SU($3$) in $4D$ adding the extra term with $\alpha=3$ gives poor estimates for all the parameters; with $\beta=5.9$, $\sigma a^2$ changes from $0.0904(70)$ to $0.08(13)$, $V_0 a$ from $0.670(21)$ to $0.72(69)$, and $\gamma$ from $-0.236(14)$ to $-0.31(97)$.

\begin{figure}[h!]
\centering
\includegraphics[width=0.95\linewidth]{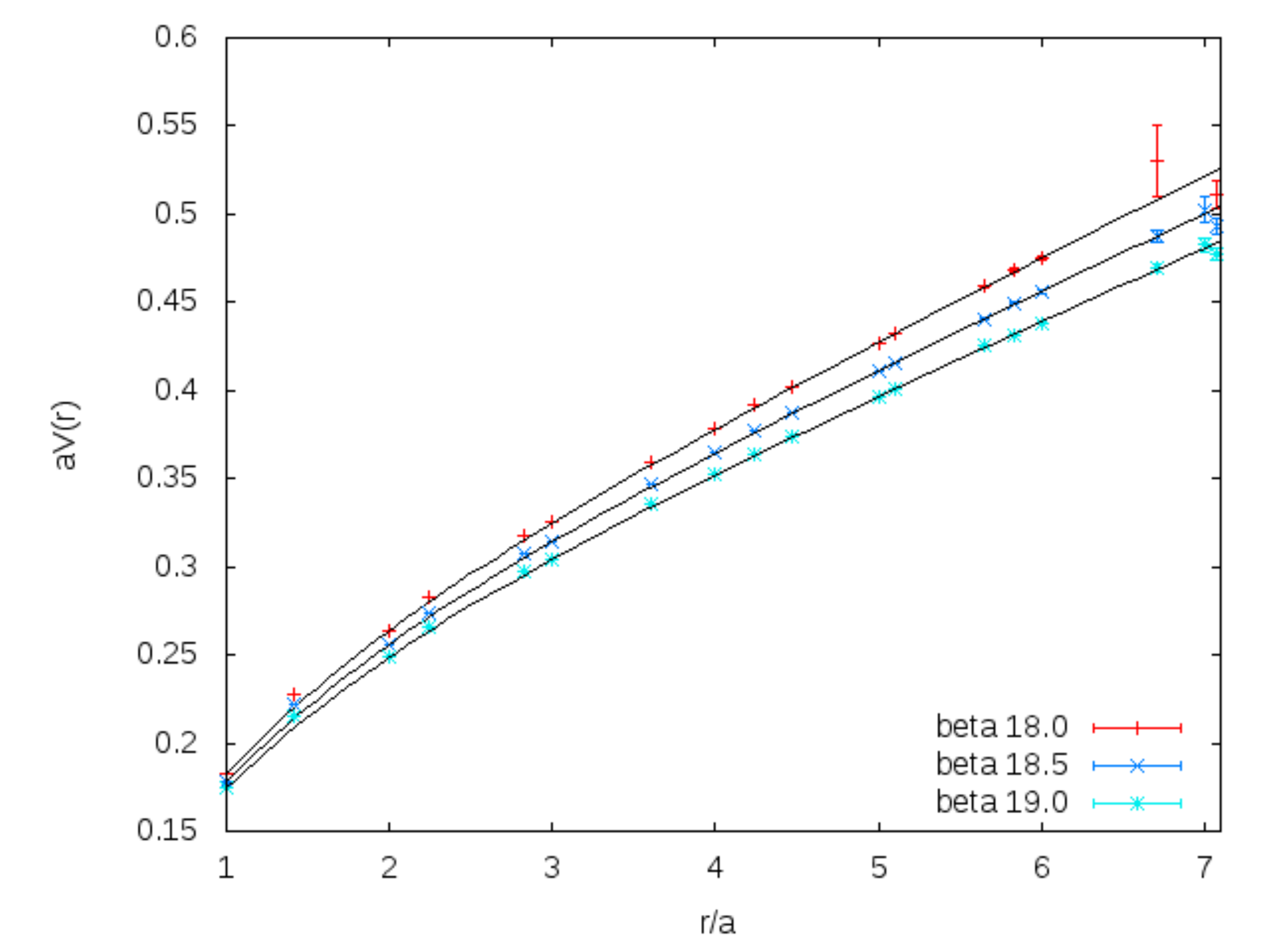}
\caption{SU($3$), in $3$ dimensions, with the Wilson action. $4 \times 10^5$ measurements, lattice size $48 \times 24^2$.}
\label{run1}
\end{figure}

\begin{figure}[h!]
\centering
\includegraphics[width=0.95\linewidth]{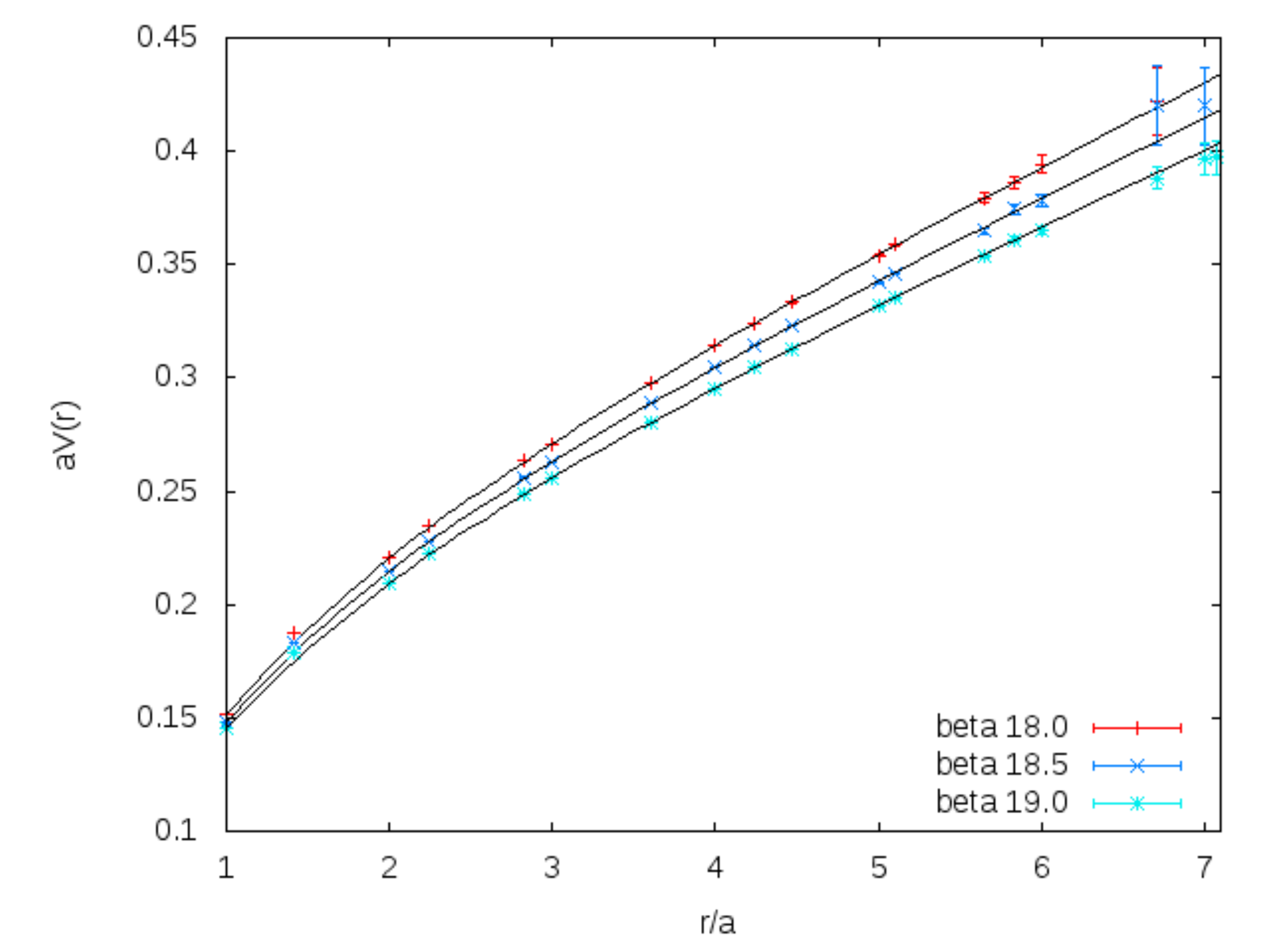}
\caption{SU($3$), in $3$ dimensions, with the improved action. $4 \times 10^5$ measurements, lattice size $48 \times 24^2$.}
\label{run2}
\end{figure}

\begin{figure}[h!]
\centering
\includegraphics[width=0.95\linewidth]{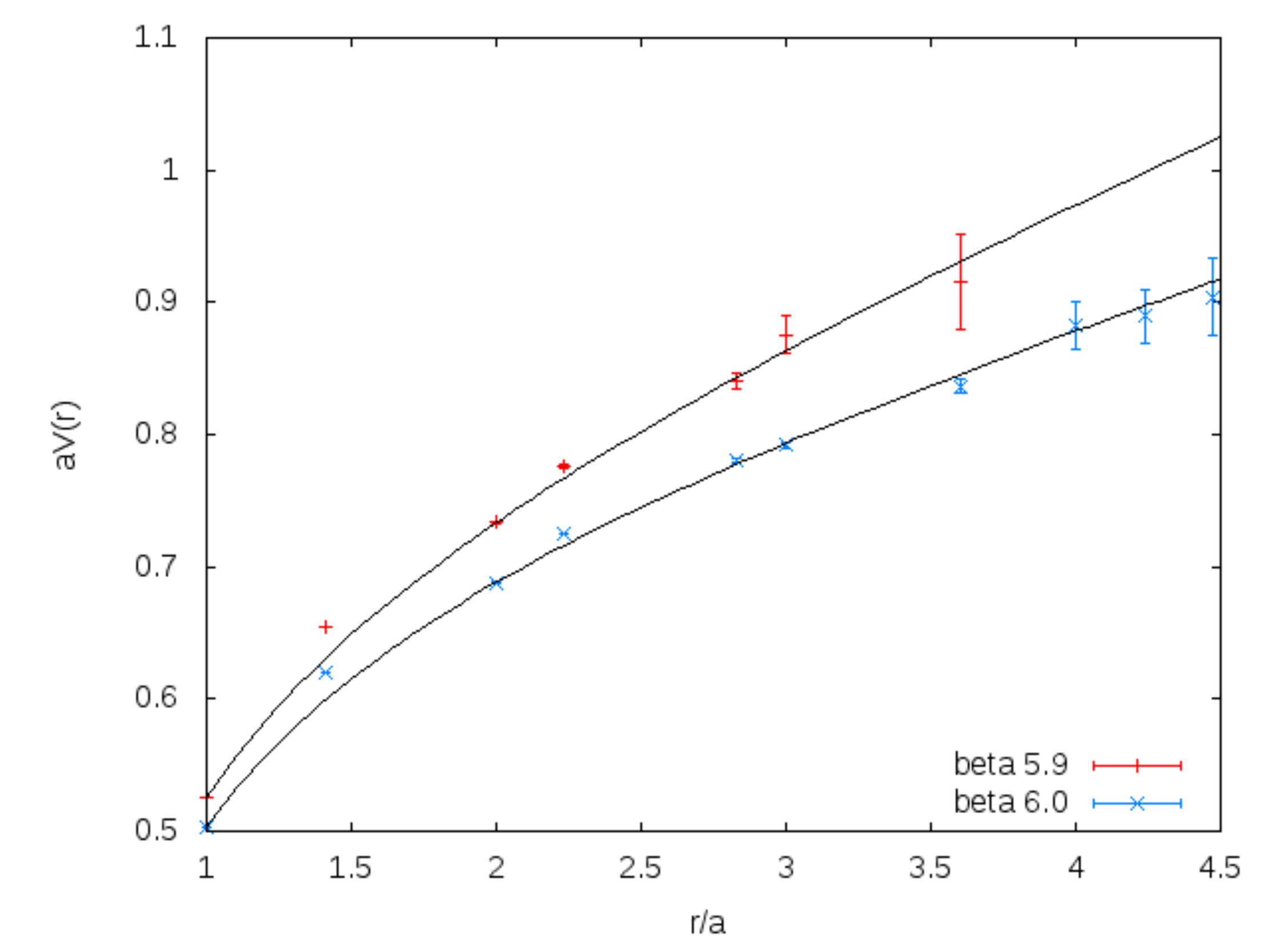}
\caption{SU($3$), in $4$ dimensions, with the Wilson action. $4 \times 10^5$ measurements, lattice size $24 \times 16^3$.}
\label{run3}
\end{figure}

\begin{figure}[h!]
\centering
\includegraphics[width=0.95\linewidth]{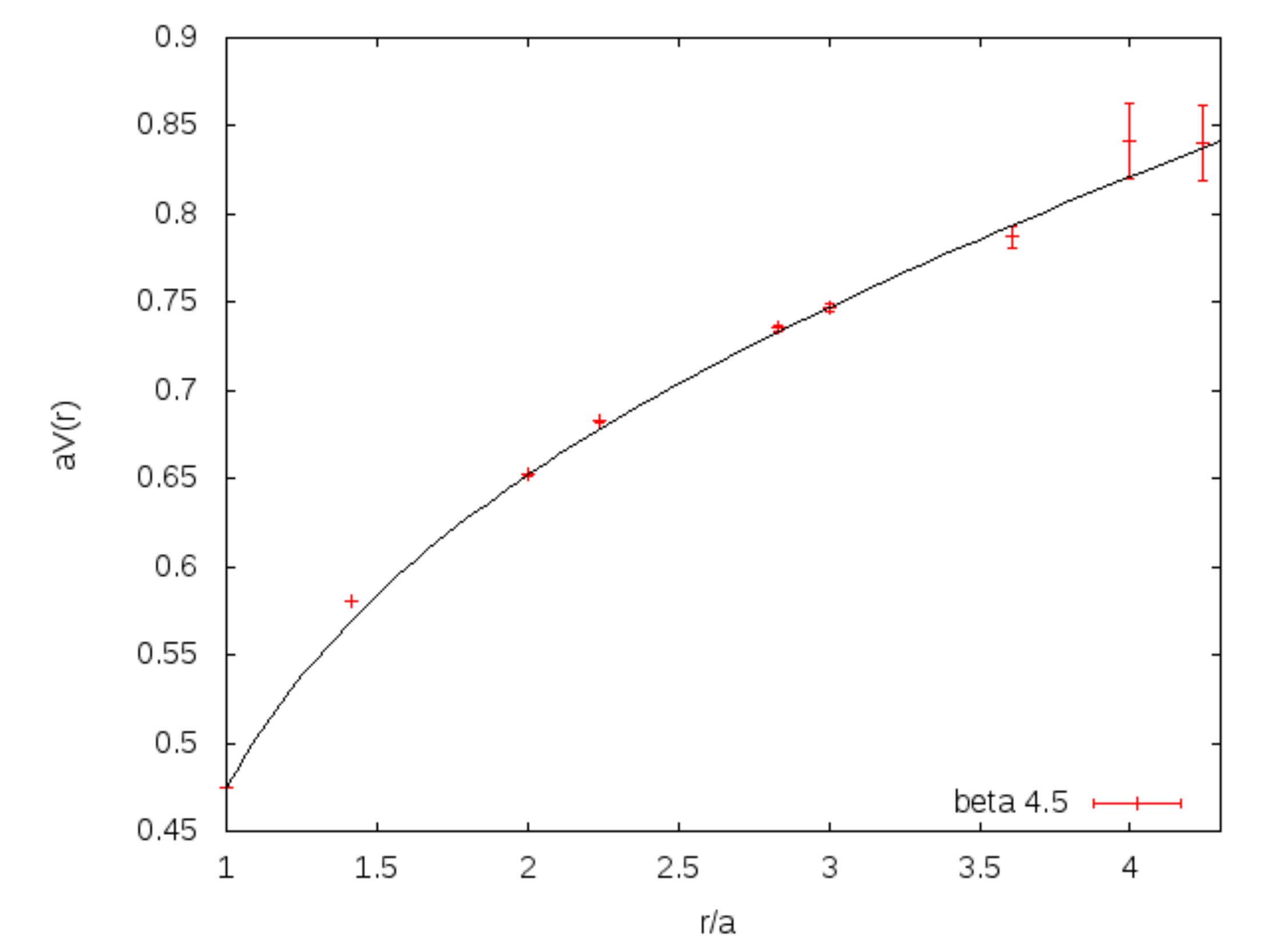}
\caption{SU($3$), in $4$ dimensions, with the improved action. $3 \times 10^5$ measurements, lattice size $24 \times 16^3$.}
\label{run4}
\end{figure}

\begin{figure}[h!]
\centering
\includegraphics[width=0.95\linewidth]{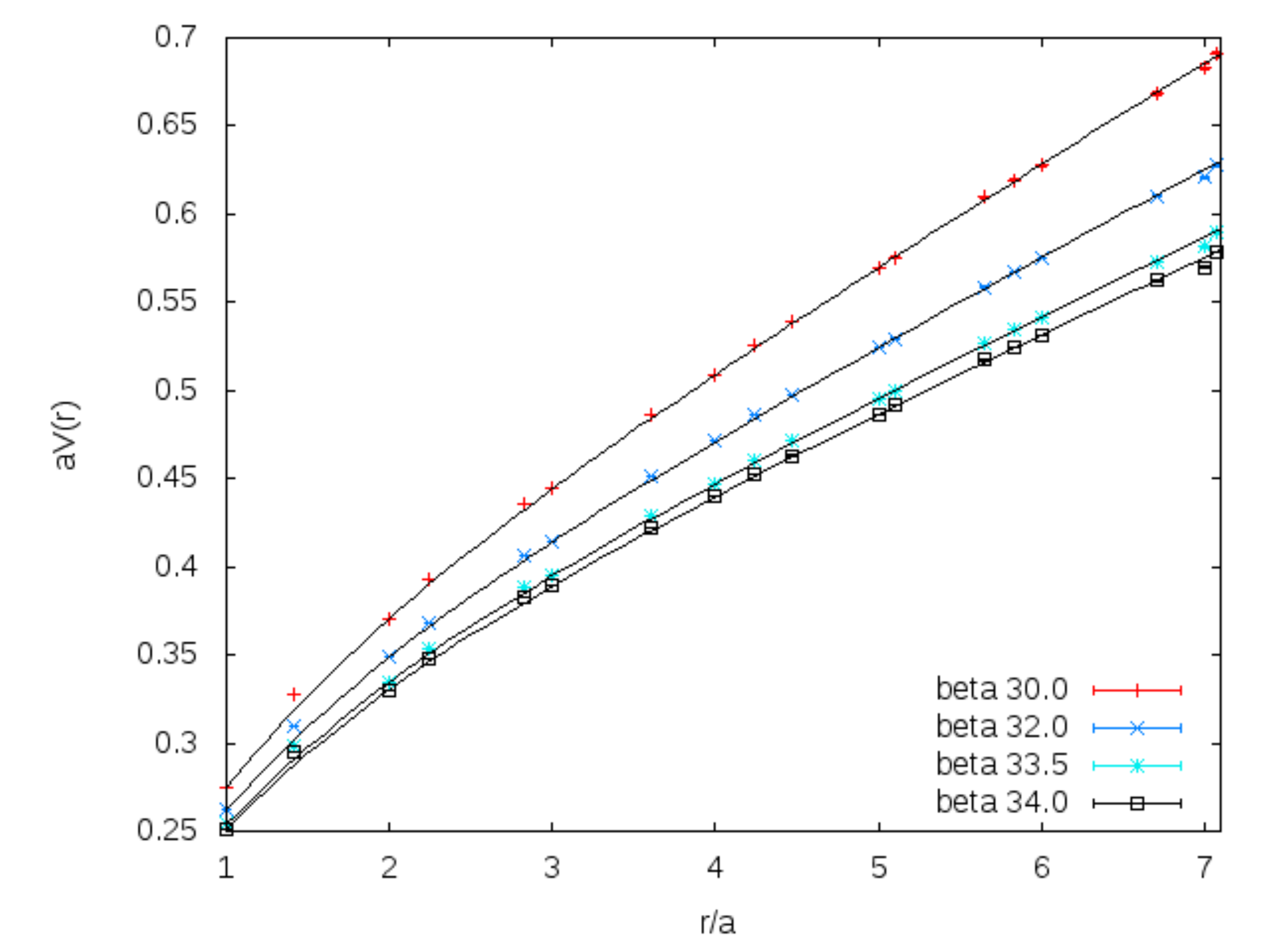}
\caption{SU($4$), in $3$ dimensions, with the Wilson action. $5 \times 10^5$ measurements, lattice size $24 \times 16^2$.}
\label{run5}
\end{figure}

\begin{figure}[h!]
\centering
\includegraphics[width=0.95\linewidth]{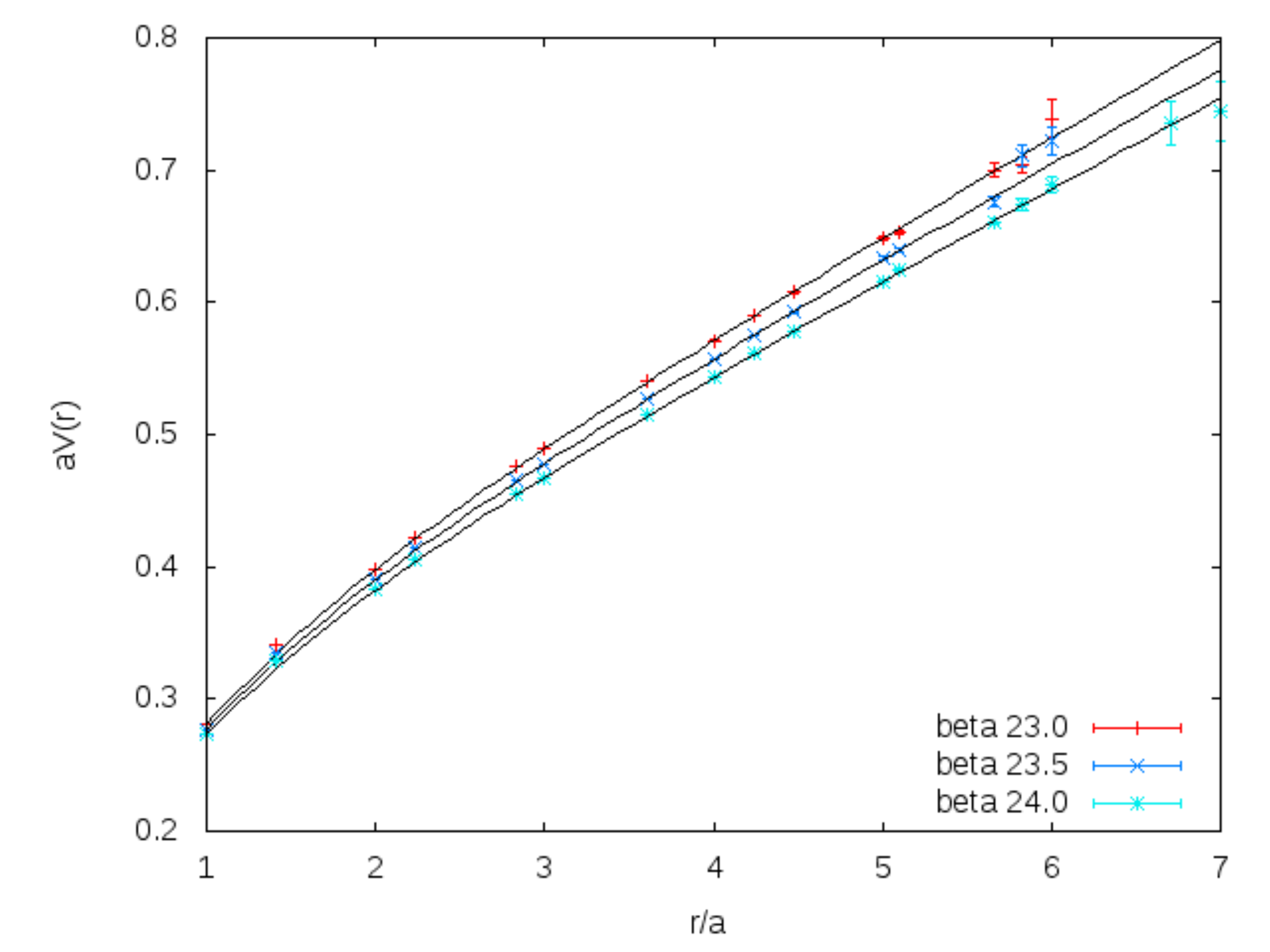}
\caption{SU($4$), in $3$ dimensions, with the improved action. $5 \times 10^5$ measurements, lattice size $24 \times 16^2$.}
\label{run6}
\end{figure}

\begin{figure}[h!]
\centering
\includegraphics[width=0.95\linewidth]{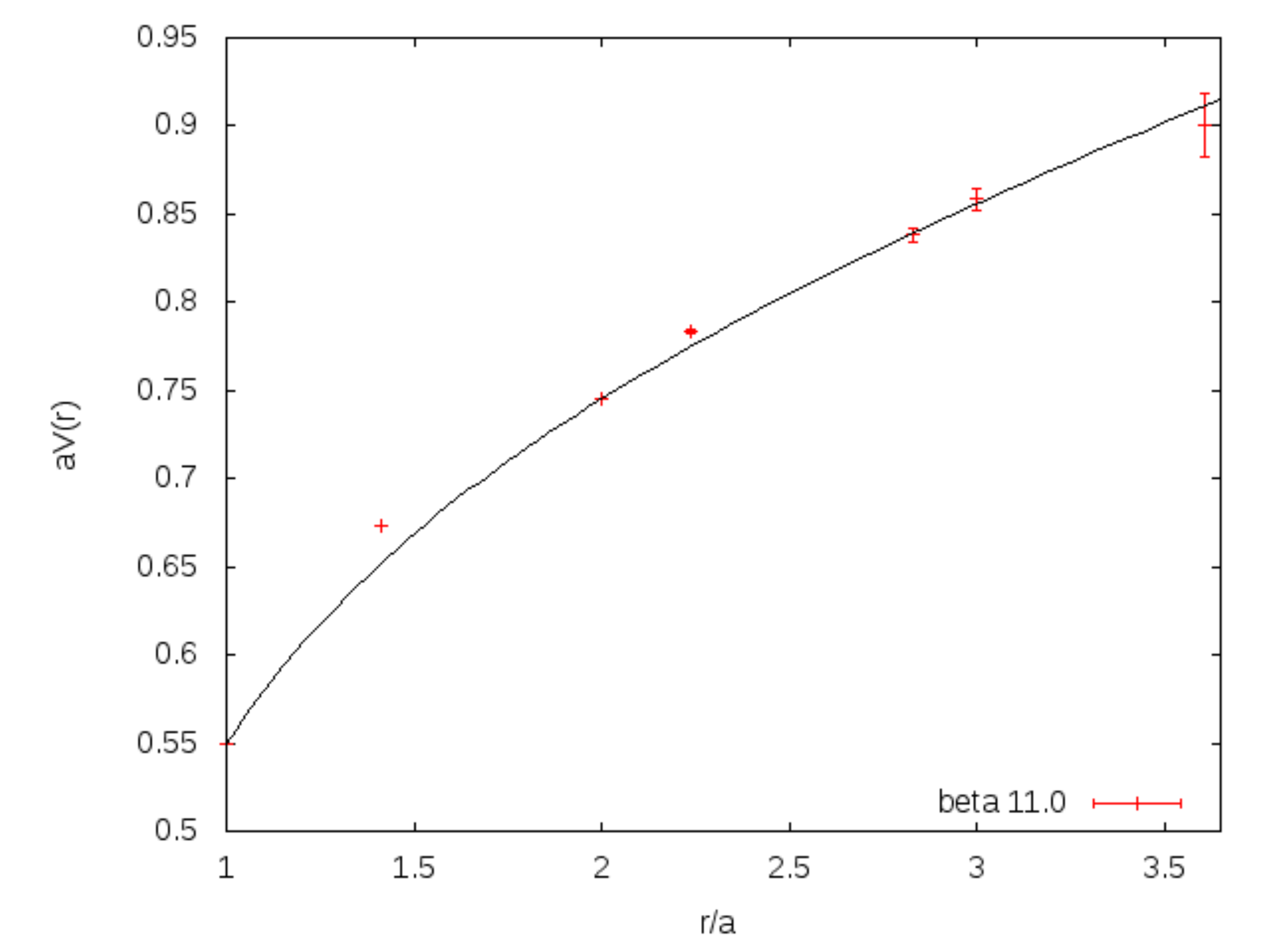}
\caption{SU($4$), in $4$ dimensions, with the Wilson action. $3.3 \times 10^5$ measurements, lattice size $24 \times 16^3$.}
\label{run7}
\end{figure}

\begin{figure}[h!]
\centering
\includegraphics[width=0.95\linewidth]{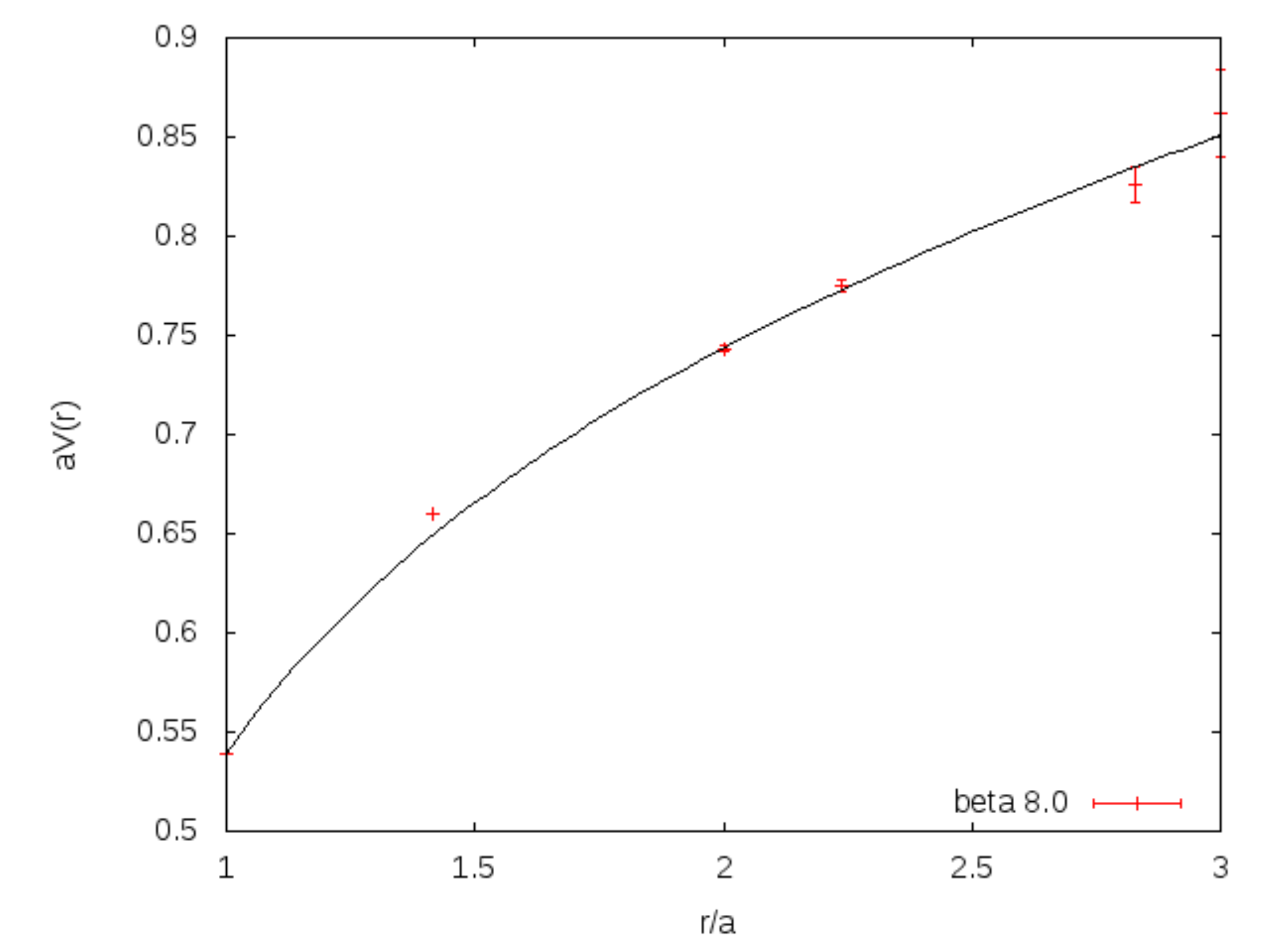}
\caption{SU($4$), in $4$ dimensions, with the improved action. $1.1 \times 10^5$ measurements, lattice size $24 \times 16^3$.}
\label{run8}
\end{figure}

The quality of the fit and, more precisely, the $\chi^2$ are heavily influenced by the choice of points that are included in the fitting process. Since the precision of our results is very high, small distance off-axis points have a huge impact; for example with SU($3$), $\beta=18.5$ in $3D$, we get $\chi^2=1.04$ when the small distance off-axis points are neglected in the fitting, but when included, the $\chi^2$ grows to a staggering $4445.21$. Obviously, this is due to systematic effects related to lattice artifacts at short distances. Already removing just a couple of these points brings the $\chi^2$ down to $107.028$. In another example, with SU($4$), $\beta=23.0$ in $3D$, we get $\chi^2=0.91$ with the small distance off-axis points neglected, and $\chi^2=3627.49 $ when they are included. For reliable results, all fits in this paper are done excluding the small distance off-axis points. The points are however included in the figures \ref{run1}-\ref{run8}.

\begin{table}[!h]
\centering
\begin{tabular}{l l l |l l l l}
$N_C$ 	& $D$ & $\beta$  & $\sigma a^2$  & $V_0 a$    & $\gamma$  & $ b/a^2 $ \\
\hline
$ 3 $ 	& $3$ & $ 18.0 $ &  $0.043927(57) $  & $0.23041(32) $  & $-0.11434(47)$ & $0.02313(23)$\\  
$   $ 	&     & $ 18.5 $ &  $0.041339(58) $  & $0.22689(33) $  & $-0.11276(48)$  & $0.02306(21) $\\
$   $ 	&     & $ 19.0 $ &  $0.039042(62) $  & $ 0.22313(36)$  & $-0.11069(53)  $  & $0.02276(23) $\\
\hline
$   $ 	& $4$ & $ 5.9 $ &  $ 0.0904(70)$  & $ 0.670(21)$  & $-0.236(14) $ & $ $\\  
$   $ 	& $ $ & $ 6.0 $ &  $0.0649(21) $  & $0.6788(63) $  & $-0.2412(42) $ & $ $ \\
\hline
$ 4 $ 	& $3$ & $ 32.0 $ &  $ 0.04716(15) $  & $ 0.31280(90) $  & $-0.1230(13) $  & $0.02528(60) $\\ 
$   $ 	& $ $ & $ 33.5 $ &  $ 0.04244(25) $  & $ 0.3070(15)$  & $-0.1210(23) $  & $0.0255(10) $\\
$   $ 	& $ $ & $ 34.0 $ &  $ 0.04106(27) $  & $ 0.3050(16)$  & $ -0.1202(25) $  & $0.0254(11) $ \\
$   $ 	& $ $ & $ 30.0 $ &  $ 0.054518(76) $  & $0.32240(44) $  & $ -0.12795(64)$   & $ 0.02581(28) $\\
\hline
$   $ 	& $4$ & $ 11.0 $ &  $0.0680(14) $  & $0.737(19) $  & $ -0.255(12)$  & $ $ \\ 

\end{tabular}
\\
\caption{Fit results with Wilson action.}
\label{wilson_table}
\end{table}

\begin{table}[!h]
\centering
\begin{tabular}{l l l |l l l l}
$N_C$ 	& $D$ & $\beta$  & $\sigma a^2$  & $V_0 a$    & $\gamma$ & $ b/a^2 $ \\
\hline
$ 3 $ 	& $3$ & $ 18.0 $ &  $ 0.034971(85)$  & $0.20028(48) $  & $-0.10488(68) $  & $0.02115(29) $ \\ 
$   $ 	& $ $ & $ 18.5 $ &  $0.03320(14) $  & $ 0.19695(76)$  & $-0.1025(11) $  & $0.02061(48) $  \\
$   $ 	& $ $ & $ 19.0 $ &  $0.031427(70) $  & $ 0.19493(40) $  & $-0.10222(58) $  & $0.02106(25) $  \\
\hline
$   $ 	& $4$ & $ 4.5 $ &  $ 0.0536(22)$  & $ 0.6686(66)$  & $ -0.2478(44)$  & $ $\\ 
\hline
$ 4 $ 	& $3$ & $23.0$ &  $ 0.07118(38)$  & $ 0.3191(21)$  & $-0.1333(31) $  & $0.0244(13) $\\  
$   $ 	& $ $ & $23.5$ &  $ 0.06852(47)$  & $0.3145(26) $  & $ -0.1292(38)$  & $0.0232(17) $\\
$   $ 	& $ $ & $24.0$ &  $ 0.06599(24)$  & $ 0.3102(13)$  & $ -0.1250(19)$  & $0.02187(85) $\\
\hline
$   $ 	& $4$ & $ 8.0 $ &  $0.058(13) $  & $ 0.771(41)$  & $ -0.290(28) $  & $ $  \\   
\end{tabular}
\\
\caption{Fit results with the improved action.}
\label{impr_table}
\end{table}

\begin{table}[!h]
\centering
\begin{tabular}{c c | c c || c c }
\multicolumn{2}{c}{ } & \multicolumn{2}{c}{Wilson action} & \multicolumn{2}{c}{Improved action}\\
\hline
$N_C$ 	& $D$ & $\beta$ & $\textrm{reduced}~\chi^2$ & $\beta$ & $\textrm{reduced}~\chi^2$ \\
\hline
$ 3 $ 	& $3$ & $ 18.0 $ & $ 1.53$ & $ 18.0 $  & $0.26 $\\  
$   $ 	&     & $ 18.5 $ & $ 2.25$ & $ 18.5 $  & $ 1.04$\\
$   $ 	&     & $ 19.0 $ & $ 3.48$ & $ 19.0 $  & $0.42 $\\
\hline
$   $ 	& $4$ & $ 5.9 $ & $0.53$ & $ 4.5 $ & $ 0.49$\\  
$   $ 	& $ $ & $ 6.0 $ & $1.21$ & $ $ & $  $\\
\hline
$ 4 $ 	& $3$ & $ 32.0 $ & $15.91$ & $23.0$ & $0.91 $\\ 
$   $ 	& $ $ & $ 33.5 $ & $ 65.21$ & $23.5$ & $2.07 $\\
$   $ 	& $ $ & $ 34.0 $ & $ 86.83$ & $24.0$ & $ 0.79$ \\
$   $ 	& $ $ & $ 30.0 $ & $ 1.81$ & $ $ & $ $\\
\hline
$   $ 	& $4$ & $ 11.0 $ & $ 0.55$ & $ 8.0 $ & $0.98$\\ 
\end{tabular}
\\
\caption{The reduced $\chi^2$ for each of the fits.}
\end{table}

\clearpage

\section{Conclusions}

In this work, we have generalized the L\"uscher and Weisz multilevel algorithm for the tree-level improved gauge action to study SU($N$) Yang-Mills theories. Applying this efficient lattice gauge theory algorithm, we computed the static quark potential in SU($3$) and SU($4$) Yang-Mills theories in $2+1$ and $3+1$ dimensions. Even with the relatively limited computational resources devoted to this first application of this algorithm, we obtained a very good level of precision. The calculated L\"uscher term was shown to satisfy the predictions of bosonic string theory, more or less equally well with the Wilson action as with the improved action. Also, we can see that the results obtained with SU($4$) agree with the results of SU($3$), and in fact, the bosonic string prediction for the L\"uscher term is met more accurately with SU($4$). This extends the previous study done in \citep{addedBonati} towards the large-$N$ limit.

Besides the calculations of the static quark potential, multilevel algorithms can be used in many other contexts, such as in the computation of glueball masses and the correlation functions related to the transport coefficients of the quark-gluon plasma. For future endeavors, in principle the multilevel algorithm for the improved gauge action could also be used in graphics processing unit (GPU) implementations. As an example of this type of application, very recently this has been done for the compact U($1$) lattice gauge theory with the standard Wilson action, in \citep{gpu}.

Another possible useful application for a high-precision multilevel algorithm with the improvement in $3D$ could be in the context of dimensionally-reduced effective theories for hot QCD, which, for example, very recently have been used to compute certain non-perturbative contributions to the jet-quenching parameter in \citep{addedBenzke, addedLaine}.

\subsection*{Acknowledgements}
I would like to acknowledge support from the Academy of Finland grant 1134018 and the Magnus Ehrnrooth Foundation. I also thank K. Kajantie, M. Panero and K. Rummukainen for valuable comments and discussions. The simulations were carried out at the Finnish IT Center for Science (CSC). 

\newpage

\bibliography{reflist}

\end{document}